\RequirePackage{fix-cm}

\documentclass[smallcondensed]{svjour3}     

\smartqed  
\usepackage{graphicx}
\usepackage{amsmath}
\usepackage{amssymb}
\usepackage{textcomp}

\journalname{Applied Physics A}
\usepackage[colorlinks]{hyperref}
\usepackage[document]{ragged2e}
\usepackage[square, comma, sort&compress, numbers]{natbib}
\usepackage{tikz}

\makeatletter
\usepackage{color}
\PassOptionsToPackage{normalem}{ulem}
\usepackage{ulem}
\providecolor{added}{rgb}{0,0,1}
\providecolor{deleted}{rgb}{1,0,0}

\begin{document}

\title {Orientation of ripples induced by ultrafast laser pulses on copper in different liquids}

\author{Stella Maragkaki         \and
  		Abdallah Elkalash         \and
        Evgeny L. Gurevich}


\institute{Stella Maragkaki \at
              Chair of Applied Laser Technology, Ruhr-Universit\"at Bochum,
Universit\"atsstra\ss e~150, 44801 Bochum, Germany \\}

\institute{Abdallah Elkalash \at
              Chair of Applied Laser Technology, Ruhr-Universit\"at Bochum,
Universit\"atsstra\ss e~150, 44801 Bochum, Germany \\
}
\institute{Evgeny L. Gurevich \at
              Chair of Applied Laser Technology, Ruhr-Universit\"at Bochum,
Universit\"atsstra\ss e~150, 44801 Bochum, Germany \\
              Tel.: +49 234/32 29891\\
%
\email{gurevich@lat.rub.de}           
}

\date{Received: date / Accepted: date}

\maketitle

\begin{abstract}

\justify
Formation of laser-induced periodic surface structures (LIPSS or ripples) was studied on a metallic surface of polished copper by using irradiation with multiple femtosecond laser pulses in different environmental conditions (air, water, ethanol and methanol). Uniform LIPSS have been achieved by controlling the peak fluence and the overlapping rate. Ripples in both orientations, perpendicular and parallel to laser polarization, were observed in all liquids simultaneously. The orientation of these ripples in the center of the ablated line
was changing with the incident light intensity. For low intensities the orientation of the ripples is perpendicular to the laser polarization, whereas for high intensities it turns parallel to it without considerable changes in the period.
Multi-directional LIPSS formation was also observed for moderate peak fluence in liquid environments.
 \justify
\end{abstract}
\section{Introduction}
\label{intro}
\justify
Laser-induced periodic surface structures (LIPSS or ripples) were reported for the first time by Birnbaum in 1965 \cite{Birnbaum}, where he revealed a system of parallel straight lines on a germanium surface upon irradiation with a ruby laser. Nowadays two types of LIPSS are studied on metal and semiconductor surfaces: (1) low spatial frequency LIPSS (LSFL), which period is comparable to the wavelength of the incident light and orientation of LIPSS perpendicular to the light polarization and (2) high spatial frequency LIPSS (HSFL), which period is several times smaller and the orientation can be either parallel or perpendicular to the polarization \cite{BonseRev}. The physical processes involved in the formation of this pattern are still not understood completely and several theories describing interaction between laser light and the solid surfaces can be discussed. 
The most probable mechanisms of the LIPSS formation involve interference of laser light on surface plasmons \cite{Sipe,Thibault1,Bonse} or hydrodynamic instabilities \cite{Costache,Varlamova,Gurevich2016}. A combination of these both mechanisms is also possible \cite{PRB_instability,Yoann2016,DifWLs}.

Analytical estimations show that the conditions for the Rayleigh-Taylor instability can be fulfilled for femtosecond laser ablation \cite{PRB_instability}. Moreover, molecular dynamic simulations of laser ablation in water demonstrate excitation of this instability and indicate its role in the formation of nanoparticles upon femtosecond laser ablation in liquids \cite{ZhigileiRTI}. In this case the deacceleration required for the instability is provided by the dense liquid environment, which stops the expansion of the molten metal surface. However, in frames of the strong explosion approximation \cite{ZeldovichRaizer}, the density of the environment plays a minor role, because the acceleration is scaled with the environmental density to the power of $1/5$. Consequently, the same mechanisms can be expected by the LIPSS formation in liquids and at atmospheric conditions. However, the effect of the liquid environment may be more complex, so it can e.g., change the resonance conditions for excitation of surface plasmons \cite{Thibault1} or broaden the spectrum of the incident beam \cite{Wittmann1996}.

The LIPSS are usually studied by ablation at atmospheric conditions in order to simplify the experiment. Several experiments on LIPSS formation on surfaces of metals and dielectrics in different liquids such as water, ethanol ($C_2H_5OH$), carbon tetrachloride ($CCl_4$), trichlortrifluorethan ($C_2Cl_3F_3$) and others, see e.g., \cite{Barmina, radu2011silicon, Thibault1} has been published recently. Experiments on LIPSS formation on silicon show that the environment changes the period of the observed pattern. For example, the LIPSS periodicity in water was five time smaller than the period in air \cite{Thibault1}, whereas in ethanol only 30\% decrease \cite{Jiao} was observed.
These studies \cite{Jiao, Thibault1} claimed that the period of LIPSS under water and ethanol is changed compared to the experiments in the air due to the shift in the plasmon resonance conditions, which depend on the refractive index of the environment.

In this paper we study LIPSS formation on the surface of copper immersed in different liquids like water, methanol and ethanol. We focus on the intensity-dependent change in the orientation of LIPSS in liquids, which is observed in liquids much stronger than in the air.

\section{Experimental Setup}
\label{sec:1}
\begin{figure}[h]
\resizebox{.9\linewidth}{!}{
\begin{tikzpicture}
   \filldraw[fill=blue!40](4.65,3.34)arc(115:65:3)--cycle;
   \filldraw[fill=blue!40](4.65,3.3)rectangle(7.2,3);
  \draw[line width=2pt](1,7.5)rectangle(4,6.5);
  \fill[fill=blue!30](4,1)rectangle(8,0);
  \draw[red,thick](4,7)--(8,7)--(6,6)--(6,0.1);
  \draw[dashed](4,1)--(8.5,1);
  \filldraw[fill=red](5,0.3)rectangle(7,0);
  \draw[dashed](7,0.3)--(8.5,0.3);
  \draw[<->](8.5,0.3)--(8.5,1)node[right,midway]{4\,mm};
  \draw[line width=2pt](4,1.5)--(4,0)--(8,0)--(8,1.5);
  \draw[line width=2pt](7.5,7.5)--(8.5,6.5);
  \draw[line width=2pt](6.5,6.5)--(5.5,5.5);
  \draw[thick,dashed](4.5,2.5)rectangle(9,8);
   \node at (2,7){\Large 1};
   \node at (8,6){\Large 2};
   \node at (7.7,3){\Large 3};
   \node at (3.5,1){\Large 4};
\end{tikzpicture}
}
\caption{Schematic representation of the experimental setup. 1 - laser; 2 - two computer-controlled mirrors of the galvo scanner; 3 - F-theta lens; 4 - Petri dish with copper sample immersed in the liquid.}
\label{setup}
\end{figure}
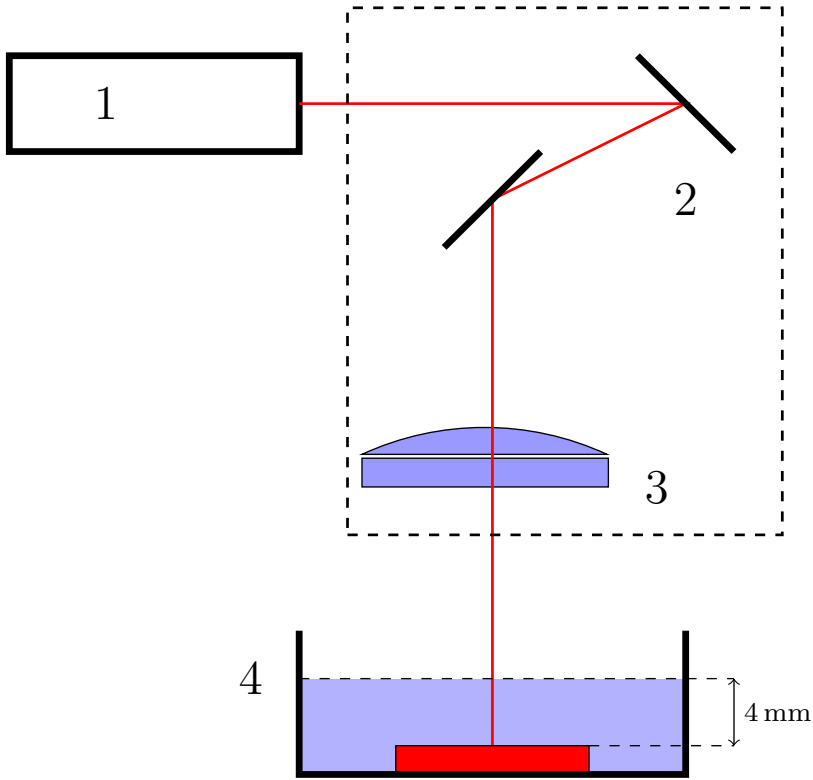

\justify
A fiber femtosecond laser ({\it Tangerine}, produced by {\it Amplitude Systems}) emitting linear polarized laser pulses delivering 1.5\,ps pulses at the wavelength $\lambda=1030$\,nm and 1\,kHz repetition rate was used for our experiments. A half lambda plate and a polarizer were used to control the laser pulse energy. Fine polished copper samples were immersed into a glass vessel so that the liquid-air interface was always 4\,mm above the sample. The copper sample was placed perpendicular to the laser beam. The laser beam was guided onto the surface through a system of galvo-mirrors and an f-theta lens, where the sample surface was positioned at the focal plane of the lens, as shown in Fig.~\ref{setup}.
\justify

\justify

Each line is written in only one run, after which a groove is left on the copper surface. The scanning direction is almost identical with the direction of the light polarisation, the deviation between them is approximately 10\textdegree{} due to the angles between the mirrors in the galvo scanner \cite{Anzolin2010}. The number of pulses per spot (N/spot) was calculated while taking into account the ablated spot diameter and the scanning speed. The peak fluence was calculated as
$\Phi_P={2E_P}/{{\pi\omega_0^2}}$ \cite{Liu}, where $\Phi_P$ is the peak fluence, $E_P$ the pulse energy and $\omega_0$ the beam waist.
The latter is estimated theoretically with $\omega_0=\frac{\lambda F}{\pi r}M^2\approx19\,$\,\textmu{}m, where $F=63\,$mm is the working distance of the lens, $r=1.15\,$mm is the beam radius at the lens and $M^2=1.05$ - the beam quality factor. We underestimate the real spot size in this way, but this estimation fits to the line width observed in our experiments.

All experiments are done at relatively high overlapping rate $OR\geq 97\%$ (N/spot$\geq 33$). Experiments at four different environmental conditions were performed (in air, water, ethanol and methanol). Immersion procedure was performed in normal atmospheric condition.

Agglomerated nanoparticles remain on the surface after each ablation experiment. In order to remove them we induce small waves on the liquid surface by a weak air flow before each next experiment. These waves cause a flow in the volume of the liquid, which removes the agglomerated particles from the sample surface. The sample position and the liquid inside the vessel remained unchanged until the end of the whole experimental procedure. A scanning electron microscop (\textit{Zeiss EVO MA 15}), with electron high tension set at 15\,kV and I\textsubscript{Probe} at 192\,pA was employed for the surface characterization.


\section{Results}
\label{sec:3}
\justify

\begin{figure}[h]
\centering
\includegraphics[width=14cm]{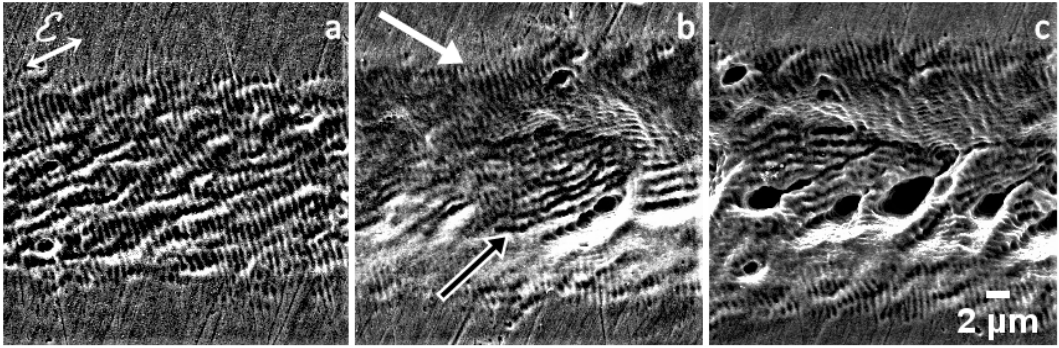}
\caption{SEM micrographs of LIPSS under 4\,mm of distilled water, where the direction of the laser polarization is indicated by the double arrow on the upper left corner, with [a] peak fluence $\Phi_P=0.36\,J/cm^2$ and $1000\,pulses/spot$, [b] $\Phi_P=0.9\,J/cm^2$, $100\,pulses/spot$, where the white arrow indicates the perpendicular LIPSS and the black the parallel LIPSS [c] $\Phi_P=0.9\,J/cm^2$, $200\,pulses/spot$.}
\label{H2O}
\end{figure}

\begin{figure}[h]
\centering
\includegraphics[width=14cm]{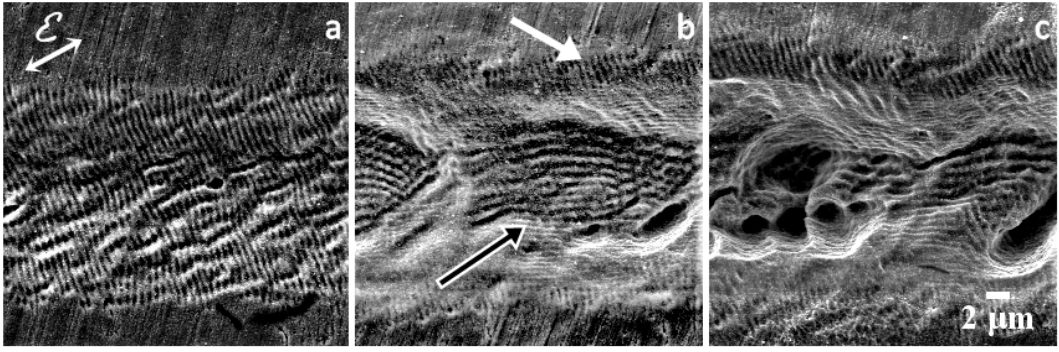}
\caption{SEM micrographs of LIPSS under 4\,mm of ethanol with [a] $\Phi_P=0.36\,J/cm^2$, $1000\,pulses/spot$, [b] $\Phi_P=0.9\,J/cm^2$, $100\,pulses/spot$, where the white arrow indicates the perpendicular ripples and the black those which are parallel to the laser polarization, [c] $\Phi_P=0.9\,J/cm^2$, $200\,pulses/spot$.}
\label{ethanol}
\end{figure}

\begin{figure}[h]
\centering
\includegraphics[width=14cm]{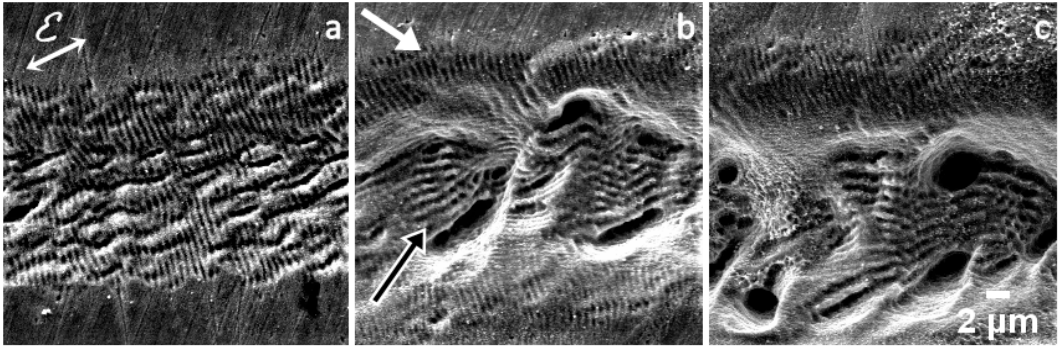}
\caption{SEM micrographs of LIPSS under 4\,mm of methanol with [a] $\Phi_P=0.36\,J/cm^2$, $1000\,pulses/spot$, [b] $\Phi_P=0.9\,J/cm^2$, $100\,pulses/spot$, where the white arrow illustrates the perpendicular ripples and the black the parallel, [c] $\Phi_P=1.8\,J/cm^2$, $100\,pulses/spot$.}
\label{methanol}
\end{figure}

\begin{figure}[h]
\centering
\includegraphics[width=9.5cm]{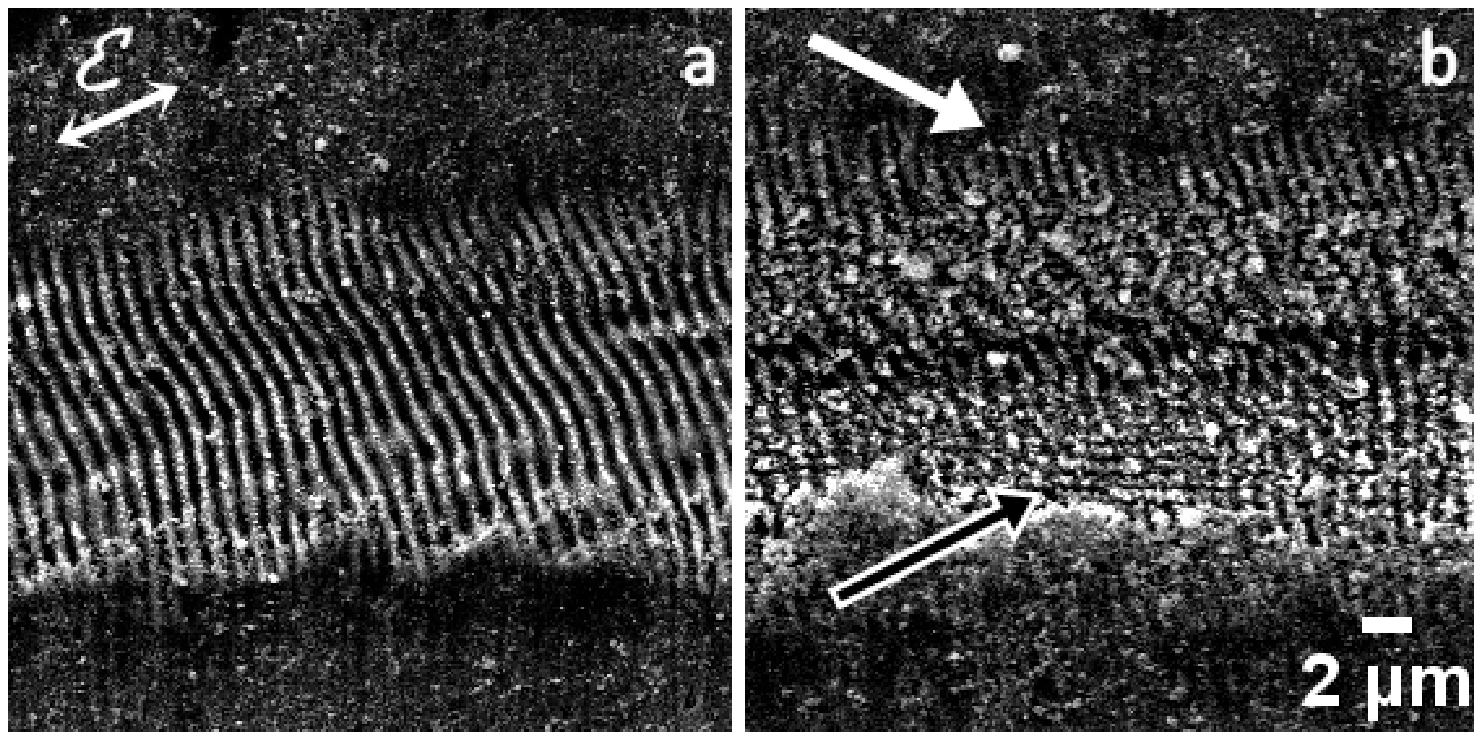}
\caption{SEM micrographs of copper irradiation in air.  [a] One-directional LIPSS at low fluences $\Phi_P=0.36\,J/cm^2$, $50\,pulses/spot$ and [b] double- directional LIPSS with $\Phi_P=0.54\,J/cm^2$, $100\,pulses/spot$, where the white arrow indicates the perpendicular periodic structures and the black the presence of parallel ripples.}
\label{air}
\end{figure}

The aim of this work is to investigate the influence of different environmental conditions on the ripples orientation. Here we refer to the orientation of the walls of the pattern as it is common in the modern literature \cite{BonseRev}, but not to the grating vector, which is perpendicular to the fringes, as in earlier publications \cite{Clark}. To assess the influence of the different liquids, the amount of each liquid above the sample surface was chosen to be the same. SEM micrographs are presented below showing that the ambient conditions have an impact on the ripples orientation (Fig.~\ref{H2O} - ~\ref{air}).  
The classical LSFL are formed with period comparable to the laser wavelength and orientation usually perpendicular to the laser polarization.
In the first set of experiments, for irradiation with low fluence - $0.36\,J/cm^2$ in water, ethanol, methanol and air (Figs. \ref{H2O}a - \ref{air}a) it was found that the orientation of LSFL on copper is always perpendicular to the polarization of the incident laser radiation,  where the period is $\Lambda$ $_{(LSFL)}$ $\approx$ 0.5 $\lambda$.

\justify

Moreover, according to the literature \cite{Hohm,Hohm2}, when higher fluences are applied, additionally to the LSFL in the center, HSFL are observed in the edge, where the light intensity is lower, with orientation parallel to the polarization.
In the second set of our experiments, a contradicting behavior is illustrated, where the ripples at higher fluences $0.54-0.9\,J/cm^2$ appear to have two different orientations, perpendicular to each other, but they all belong to LSFL. In (Figs. \ref{H2O}b - \ref{air}b) the white arrows indicate the perpendicular ripples and the black arrows the parallel ripples in the center. Thus, at the edges the ripples are perpendicular to the laser polarization, while at the center their orientation surprisingly becomes parallel. This effect is observed in all ambient conditions, although much more intense and well confined in liquids and hardly detectable in the air.

As a third step, we increased further the fluence and/or number of pulses. Obviously the formation of bubbles and inhomogeneous material ablation start, although the previously-described behavior is still presented as shown in (Figs. \ref{H2O}c-\ref{methanol}c).

\justify
\begin{figure}[h]
\centering
\includegraphics[width=12cm]{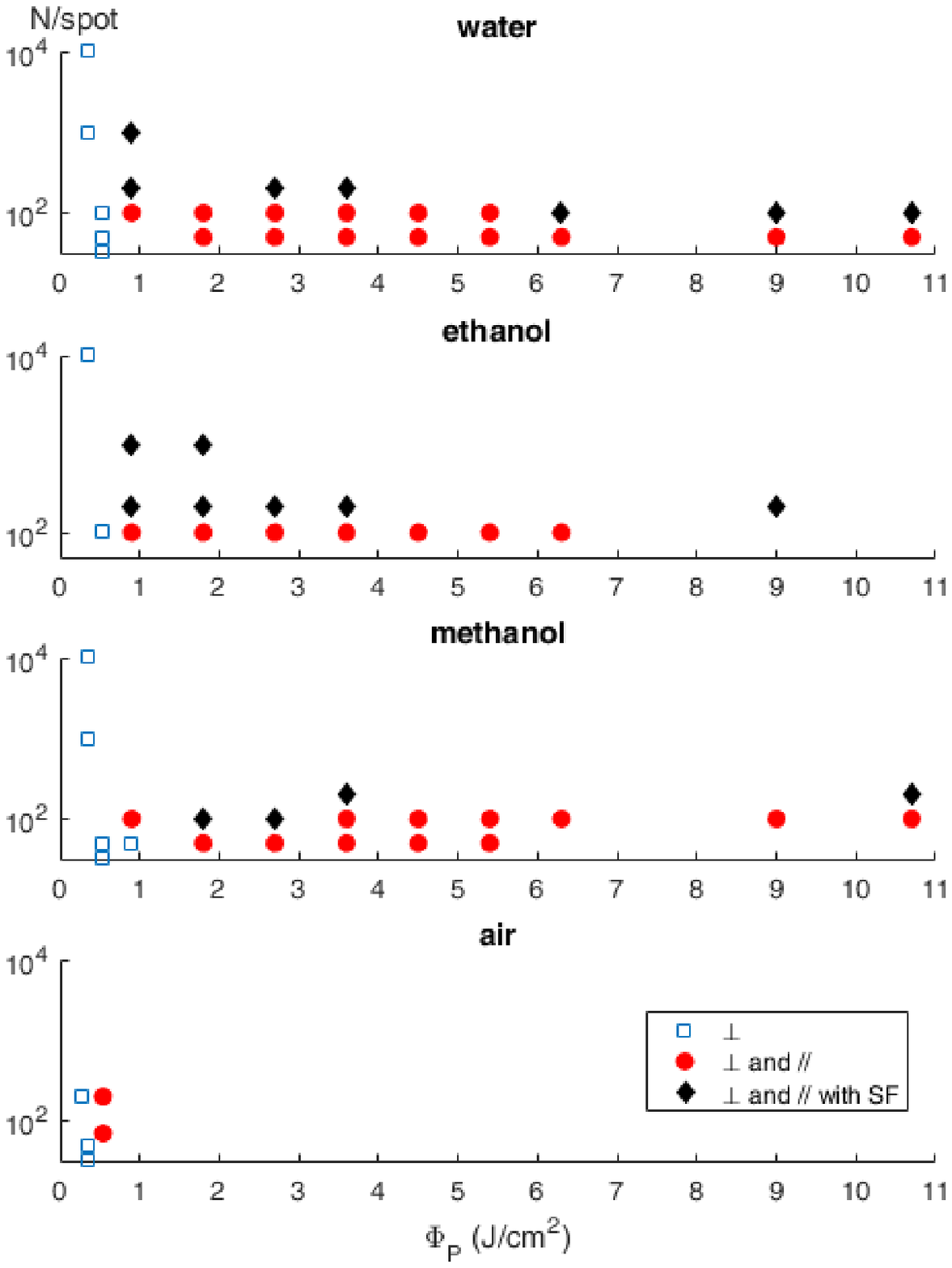}
\caption{The range of fluences and number of pulses per spot at which one-directional and two-directional LIPSS appear in all four environmental conditions. The empty square corresponds to the grooves with LIPSS perpendicular to the laser polarization, the filled circles to the LIPSS with both horizontal and perpendicular polarazation while the filled diamonds represent the two-directional LIPSS while strong influence of self-focusing is observed.}
\label{plot}
\end{figure}

\justify
\begin{figure}[h]
\centering
\includegraphics[width=2.85cm]{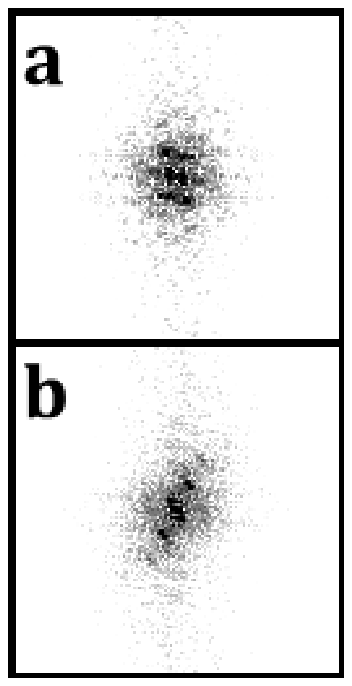}
\includegraphics[width=5.7cm]{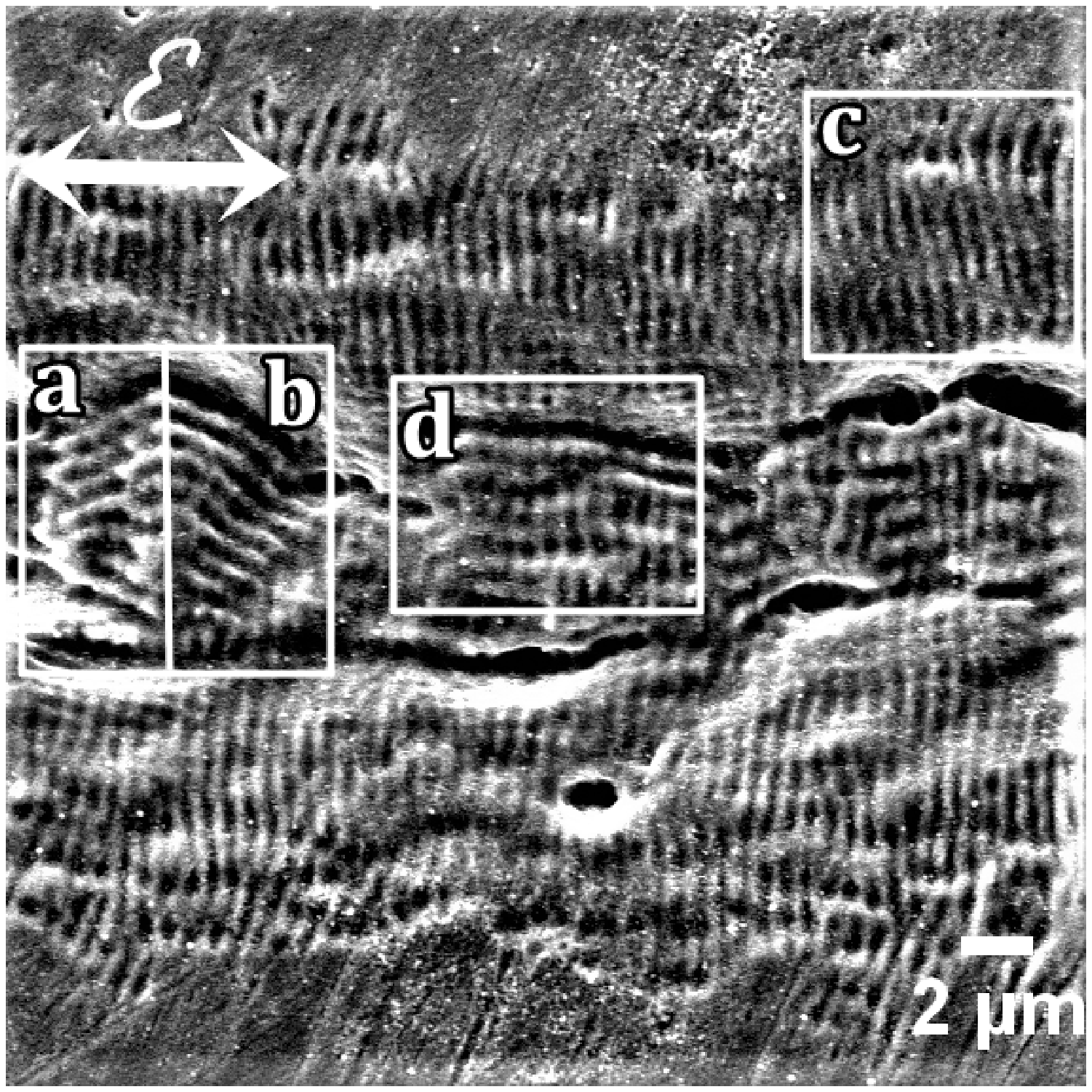}
\includegraphics[width=2.85cm]{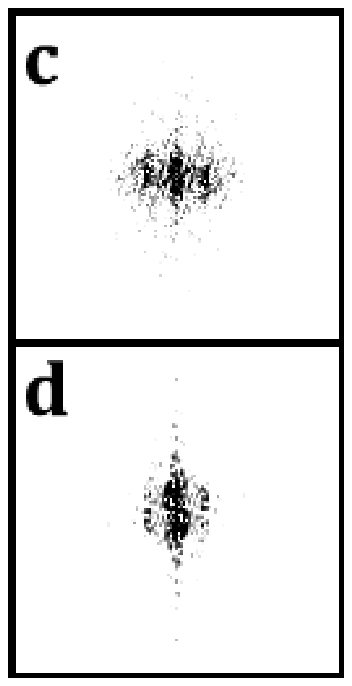}
\caption{SEM micrograph of two-directional LIPSS on copper in ethanol solution with the laser polarization parallel to the scanning direction, peak fluence $\Phi_P=0.9\,J/cm^2$ and $N/spot=200$. (a-d) The 2D-FFT from four different areas of the presented SEM image }
\label{Eparallel}
\end{figure}

\justify
\begin{table}[h]
\caption{One-directional LIPSS parameters on copper surface from (Figs. \ref{H2O}a-\ref{air}a) in different environments with fix $\tau$ =1.5\, ps, $\lambda_0$=1030\,nm, $\nu$=1\,kHz, ($n$) and ($n_2$) linear and nonlinear refractive index \cite{Elsevier}, (P$_{s\!f}$) threshold power for self-focusing, ($\Lambda$) LIPSS period, ($\Phi$\textsubscript{p}) peak fluence,  ($P_p$) peak power, ($N/spot$) number of pulses per spot, ($\perp$) and ($\parallel$): orientation of LIPSS perpendicular or parallel to the laser polarization}
\label{tab:1}       

\begin{tabular}{llllllllc}
\hline
medium& $n$ & $n_2$\,[cm$^2$W$^{-1}$] & P$_{s\!f}$\,[MW] & $\Lambda$[\,nm]& $\Phi$\textsubscript{p}[\,J/cm\textsuperscript{2}]& P\textsubscript{p}\,[MW] &N/spot&($\perp$)/($\parallel$)\\
\noalign{\smallskip}\hline\noalign{\smallskip}
air     &1    & 5.0$\,10^{-19}$ & 3200   &  895$\pm$26& 0.36 & 1.3 & 50&($\perp$)\\
water   &1.33 & 4.1$\,10^{-16}$ & 2.9 & 640$\pm$20& 0.36   & 1.3    & 1000& ($\perp$)\\
ethanol &1.36 & 7.7$\,10^{-16}$ & 1.5 & 580$\pm$27& 0.36 &  1.3 &1000& ($\perp$)\\

methanol &1.33 & 6.9$\,10^{-16}$ & 1.7 & 602$\pm$22& 0.36 & 1.3 &1000&($\perp$)\\
\hline
\end{tabular}
\end{table}
\justify

\begin{table}[h]
\caption{Two-directional LIPSS parameters on copper surface from (Figs. \ref{H2O}b-\ref{methanol}b) in different environments with ($\Lambda$) LIPSS period, ($\Phi$\textsubscript{p}) peak fluence, ($P_p$) peak power, ($N/spot$) number of pulses per spot, ($\perp$) and ($\parallel$): orientation of LIPSS perpendicular or parallel to laser polarization}
\label{tab:2}       
\begin{tabular}{lllllll}
\hline
medium&  $\Lambda$(\,nm)& $\Phi$\textsubscript{p}(\,J/cm\textsuperscript{2})& P\textsubscript{p}\,(MW)  &$N/spot$&($\perp$)/($\parallel$)\\
\noalign{\smallskip}\hline\noalign{\smallskip}
water         &620$\pm$37     &0.9               &  3.3        &100    &($\perp$)\\
$\sim$  &760$\pm$128    &$\sim$            & $\sim$        &$\sim$  &($\parallel$)\\
ethanol        &575$\pm$56    &0.9              &  3.3      &100 &($\perp$)\\
$\sim$    &640$\pm$87     &$\sim$            & $\sim$      & $\sim$ &($\parallel$)\\
methanol        &630$\pm$30     &0.9               &   3.3       &100 &($\perp$)\\
$\sim$    &790$\pm$205     &$\sim$            & $\sim$       &$\sim$  &($\parallel$)\\
\hline
\end{tabular}
\end{table}
\justify

The experimental results are summarized in the tables(~\ref{tab:1}-\ref{tab:2}). In  table~\ref{tab:1} the optical constants of the liquids as well as the critical power for self-focusing, estimated as $P_{s\!f}\approx0.15\frac{\lambda^2}{ nn_2}$ for a Gaussian beam \cite{Bauerle}, can be found.

\section{Discussion}

We observe that at high intensities of the incident laser radiation, the orientation of LIPSS in liquids (water, ethanol, methanol) changes and becomes parallel to the polarisation of light. We cannot give any final explanation of this effect but can provide some ideas about the background physics, based on the following observations:
\begin{enumerate}
\item The LIPSS orientation changes in liquids whereas in the air at comparable intensities of the incident light this effect is less pronounced. This indicates that the LIPSS reorientation is caused (or strongly increased) by the interaction with the liquid environment. 
\item For high intensities of the incident light in liquids, the orientation of LIPSS is changed only in the middle of the line, where the local intensity of the Gaussian beam is at maximum. This indicates that the effect depends on the local beam intensity, and there is a threshold, at which the orientation change takes place. For lower intensities the orientation of LIPSS remains always perpendicular to the light palarisation.
\item The periods of the perpendicular and parallel LIPSS are comparable. Hence, this effect cannot be explained by the transition from LSFL to HSFL, because the period of these structures differs by almost one order of magnitude \cite{BonseRev}.
\end{enumerate}

The following mechanisms can play role in the rotation of the LIPSS orientation:

\textit{Surface deacceleration caused by the liquid.} As suggested in \cite{ZhigileiRTI} if metals are ablated in a liquid environment, the melt is stopped due to interaction with the liquid and this deacceleration can be sufficient to develop the Rayleigh-Taylor instability, which can cause the LIPSS \cite{Gurevich2016}. As mentioned before \cite{ZeldovichRaizer}, the acceleration of the melt surface should be scaled as the fifth-order root of the environment density, so one can expect slightly different conditions for the stopping dynamics (and hence for the development of the Rayleigh-Taylor instability) in air and in liquids, but no considerable difference between different liquids. However this mechanism cannot directly explain the change in the LIPSS orientation.  

\textit{Formation of bubbles.} Theoretically bubbles can scatter the incident light and distort the wave front, but in our experiments this effect is reduced due to low repetition rate of the laser pulses (in all experiments $\nu=10^3\,$Hz) and low peak fluence. The observed groove widths at high and low fluences (see e.g., Figs.~\ref{H2O}-\ref{methanol}) are very similar, which indicates minor effect of the light scattering on the bubbles. Moreover, the grooves are continuous and never interrupted, which happens as a big bubble sticks to the surface \cite{gramNPs}. There are also some indications in the figure~\ref{air}b that `two-directional' LIPSS may be observed in the air, where no bubble formation is possible.

\textit{Self-focusing.} The peak powers, at which the LIPSS orientation changes are reasonably close to the threshold power for self-focusing as shown in tables(\ref{tab:1}-\ref{tab:2}). Moreover, the depth of the ablated line becomes inhomogeneous with deeper channels as the intensities and pulse overlaps become very high Figs. \ref{H2O}c-\ref{methanol}c. On the one hand, these deep channels cannot be explained by the bubble formation because a bubble would act as a defocusing lens due to lower refractive index of the vapor. On the other hand, the self-focusing must be independent on the overlap (except of change in $n_2$ with the temperature), whereas the bubble formation is strongly facilitated at higher overlaps. The self-focusing is known to modify the wavelength of light and generate plasma in the environment, but it is not clear how exactly it can turn the polarisation of the incident light on 90\textdegree.

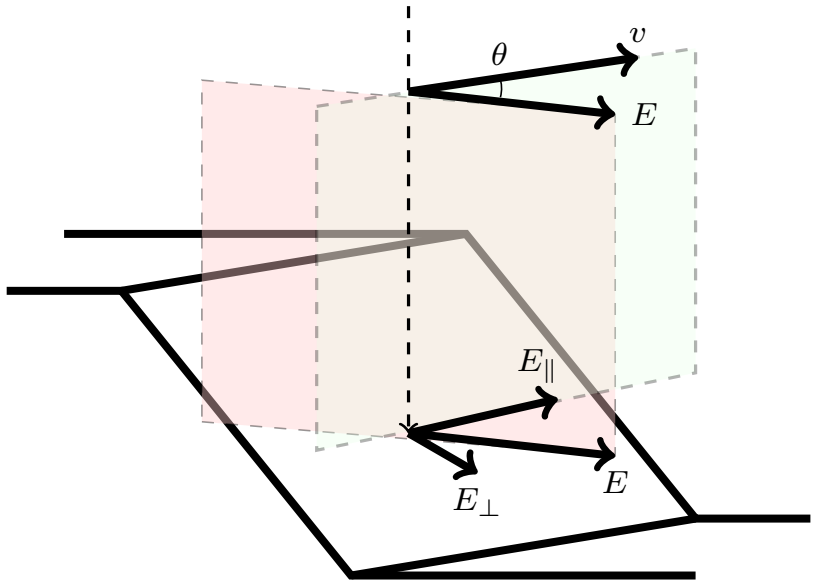
\begin{figure}[h]
\resizebox{.9\linewidth}{!}{
\begin{tikzpicture}
  \draw[line width=2pt](0.5,3.5)--(4,3.5)--(6,1)--(7,1);
  \draw[line width=2pt](0,3)--(1,3)--(3,0.5)--(6,0.5);
  \draw[line width=2pt]((3,0.5)--(6,1);
  \draw[line width=2pt](1,3)--(4,3.5);
  \filldraw[dashed,fill=red!20,opacity=0.4](1.7,1.85)--(5.3,1.55)--(5.3,4.55)--(1.7,4.85)--cycle;
  \filldraw[thick,dashed,fill=green!10,opacity=0.3](2.7,4.62)--(6,5.13)--(6,2.28)--(2.7,1.60)--cycle;
  \draw[line width=2pt,->](3.5,1.75)--(5.3,1.55)node[below]{$E$};
  \draw[line width=2pt,->](3.5,1.75)--(4.8,2.05)node[above]{\textbf{$E_\parallel\quad$}};
  \draw[line width=2pt,->](3.5,1.75)--(4.1,1.4)node[below]{$E_\perp$};
  \draw[line width=2pt,->](3.5,4.75)--(5.3,4.55)node[right]{$E$};
  \draw[line width=2pt,->](3.5,4.75)--(5.5,5.05)node[above]{$v$};
   \draw[->,thick,dashed](3.5,5.5)--(3.5,1.75);
   \draw(4.3,4.7)arc(-15:20:0.3cm)node[above]{$\theta$};
\end{tikzpicture}
}
\caption{Schematic representation of the incident electric field and the groove ablated by the laser. The angle between the polarisation and the scanning direction is $\theta$. On the plane surface the LIPSS are perpenducular to the direction of the electric field, but on the sloped surface $E_\parallel=E_s$ and $E_\perp=E_p$, thus the p-polarisation, even small, place key role in the LIPSS formation.}
\label{slope}
\end{figure}

\textit{Surface morphology.} At high intensities of the incident light, the sample is ablated and deep grooves are formed, so that the following laser pulses shine on a slightly tilted surface of the groove side walls. 
The direction of the scan (and hence, of the groove) is close, but not absolutely equal to the direction of the polarisation of the light due to the polarisation rotation in the scanner \cite{Anzolin2010}. Thus, there is a small p-component of the incident electric field on the side surface of the groove, see figure~\ref{slope},  which can either excite plasmons like it happens in the Kretschmann configuration, or increase the absorption for the p-polarisation of the incident light ($E_\perp$ in figure~\ref{slope}), or provide necessary conditions for the erosion/smoothing model.
On the plane surface the conditions for the Kretschmann mechanism of the plasmon excitation are not fulfilled \cite{SPRPF} and the absorption is just about 3-4\%, but on the sloped wall of the groove the light shines onto the metal at a higher angle and the absorption of the p-component of the polarisation increases. We notice that the conditions for the surface plasmon resonance are still not fulfilled in this case, unless a thin dielectric layer (e.g. vapor) with a real refractive index smaller than that of the liquid appears on the interface \cite{Otto1968}.

\begin{figure}[h]
\centering
\includegraphics[width=8cm]{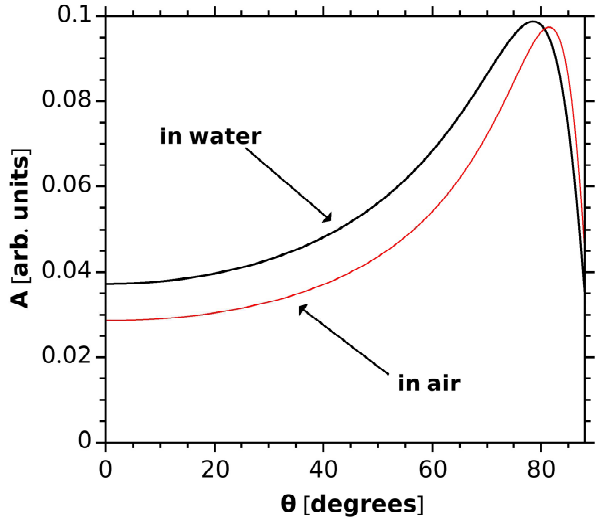}
\caption{Calculated absorption of copper for different incident angles in water (thick black curve) and in the air (thin red curve) in dependence on the incident angle $\theta$ (which corresponds to the slope angle of the side walls of the groove).}
\label{Abs}
\end{figure}

But also without the plasmon resonance, the absorption of the p-polarisation increases, see figure~\ref{Abs}. The absorption is calculated as $A=1-R$ with $R$ - reflectivity, calculated by means of {\it Winspall} software \cite{WinSpall}. In frame of this approach, the orientation of LIPSS must be also defined by the slope of the surface, which agrees with our observations. Profilometer measurements revealed a slope of the side walls of the groove in the range of $20-67$\textdegree. Figure~\ref{Abs} demonstrates, that the increase in the absorption of the p-component of the incident wave, though less pronounced, can be observed in the air also. However liquids with higher refractive index intensify this effect for the groove angles $\theta\lesssim 80$\textdegree.


\justify

\textit{Surface charging.} The importance of the surface morphology for multi-pulse LIPSS formation was also discussed in \cite{Varlamova} in the context of competition between diffusion and desorption of atoms from the charged surface upon femtosecond laser ablation. This model was further developed in \cite{AIPconf,VarlamovaBook}, where the Kuramoto-Sivashinsky-like equation was derived for the surface profile evolution. This model predicts that the LIPSS orientation is changed with the incident angle if the light energy is deposited asymmetric in the plane of the surface. Although the physical reason of the charging in our experiments remains unclear due to high electric conductivity of copper, it is possible that the liquids chemically react with the copper target. One can assume that this reaction passivates the surface and reduces its electrical conductivity. Laser-induced chemical modification was reported for iron in \cite{SCHAAF2002,KANITZ2016}.

%
%
\section{Conclusions}
\label{sec:4}
We demonstrated experimentally, that the orientation of LIPSS on copper ablated with multiple femtosecond laser pulses can deviate from the polarisation of the incident light beam. This effect has been observed in all tested liquids, which are water, ethanol and methanol in a broad range of fluences $\Phi_P\approx 1-10\,J/cm^2$. In the air less noticeable change in the LIPSS orientation has been detected only at intensities $\Phi_P\approx 0.54\,J/cm^2$ and number of pulses $N=100-200\,$pulses per site. Although the complete theory of the observed phenomenon is not clear, different possible mechanism have been discussed. Most probably, the slope of the walls of the grooves, which are produced upon laser ablation, changes the local incident angle of the light, so that the absorption of the p-polarisation component increases. From this point of view, liquid environment with high refractive index increases the absorption of this component of the incident electric field. It is also possible that the liquid facilitates the surface charging, which triggers the change in the ripples orientation. The self-focusing and bubble formation may contribute to the inhomogeneous groove profile, which is responsible for the state with mixed LIPSS orientation.

\section*{Author contribution statement.} S. Maragkaki: conception of the experiments and collection of the data. A. Elkalash: first observed the change in the LIPSS orientation. E. L Gurevich: data analysis and interpretation.



\end{document}